\documentstyle[aps,epsfig,psfig,preprint,12pt]{revtex}
\begin{document}

\baselineskip=24pt
\draft

\title{Improved approach to the heavy-to-light form factors in the light-cone QCD sum
rules \thanks{ This work is in part supported by the National Science
Foundation of China }}

\author{Tao Huang$^{1,2}$, \thanks{The mailing address:
huangt@alpha02.ihep.ac.cn}
Zuo-Hong Li$^{1,3}$ and Xiang-Yao Wu$^{2}$ }

\address{ {\footnotesize 1. CCAST (World Laboratory), P.O. Box 8730,
Beijing 100080, China } \\
{\footnotesize 2. Institute of High Energy Physics, P.O. Box 918(4),
Beijing 100039, China } \thanks{Mailing address} \\
{\footnotesize 3. Department of Physics,
Peking University, Beijing 100871, People's Republic of China} }
\maketitle

\renewcommand{\thesection}{Sec. \Roman{section}} \topmargin 10pt
\renewcommand{\thesubsection}{ \arabic{subsection}} \topmargin 10pt

{\vskip 5mm
\begin {minipage}{140mm}
\centerline {\bf Abstract}
\indent
A systematic analysis shows that the main uncertainties in the
form factors are due to the twist-3 wave functions of the light mesons
in the light-cone QCD sum rules. We propose an improved approach, in
which the twist-3 wave functions don't make any contribution and 
therefore the possible pollution by them can be avoided, to
re-examine $B \rightarrow \pi$ semileptonic form factors. Also, a
comparison between the previous and our results from the light-cone QCD
sum rules is made. Our method will be beneficial to the precise 
extracting of
$\mid{V_{ub}}\mid$ from the experimental data on the processes  
$B \rightarrow \pi \ell \widetilde{\nu_\ell}$.

\end {minipage}
}

\vspace*{2cm}
{\bf PACS numbers 13.20.He 11.55.Hx}

\newpage
\section * {I. INTRODUCTION}

Heavy-to-light exclusive decays are an important ground understanding and
testing the standard model (SM), since they can provide signal of
CP-violation phenomena and, perhaps, a window into a new physics beyond the
SM, thereby, it is of crucial interest to make a reliable prediction of these
exclusive processes. We have to
confront calculations of the hadronic matrix elements, in which all the
long-distance QCD dynamics is included. At present, an exact estimate of
them is impossible, to the present knowledge of QCD, from the first
principle, and one must resort to phenomenological approaches. Usually,
some of the methods used widely are QCD sum rules, chiral
perturbation theory (CHPT), heavy quark effective theory (HQET) and quark
model.
Each of them has advantages and disadvantages. For example, CHPT and
HQET, as two effective theories at low energy, can describe
light-to-light and heavy-to-heavy exclusive transition, respectively; but
they are not suitable for a study on heavy-to-light processes. It is 
more complicated to calculate the heavy-to-light decays. In this case,
QCD sum rule method was adapted extensively. However, it keeps some
questions left. The most striking problem is that the resulting sum rules
for form factors behave very badly in the heavy quark limit
$m_q\rightarrow\infty$. The reason is that in the operator product
expansion (OPE) at the small distance $x\approx 0$, one omits the
effect of the finite correlation length between
the quarks in the physical vacuum. In order to overcome the defect,
light-cone QCD sum rule approach is developed in \cite{s1} and is 
regarded as an advanced tool to deal with heavy-to-light exclusive
processes.
Especially, the results consistent with the physical picture can be
driven in this framework. Compared with the traditional QCD sum
rules, light-cone QCD sum rule approach is of the following different
points: the OPE is carried out near the light cone $x^2\approx 0$, instead
of at the short distance $x\approx 0$; the nonperturbative dynamics is
parametrized as so-called light-cone wave functions, instead of the vacuum
condensates. There are a lot of applications of light cone QCD sum rules
in literature. For a detailed description of this method, see \cite{s2}.

At first sight, the heavy-to-light decays
can be calculated by perturbative QCD (PQCD) due to the hard
gluon exchange (the large $Q^2$ transfer). A detailed analysis \cite{s3}
shows that the reliable PQCD calculation depends on whether the
singularities can be eliminated or suppressed by the distribution
amplitude. The singularities include on-shell gluon, on-shell light 
quark and on shell heavy quark. Carlson et al. \cite{s4} argued that
the on shell heavy quark in the hard scattering travels only a short 
distance and the factorization of the formalism still hold. Even that,
one can find that the reliable PQCD contribution may dominate only as
$m_b$ takes some special values and $\phi_\pi = \phi_\pi^{as}$ \cite{s3}.
In order
to make PQCD applicable, Ref. \cite{s5} adapt the modified
hard-scattering approach to the case of the heavy-light form factor by
a resummation of Sudakov logarithms which may suppress the soft
contribution beyond naive power counting. However, this approach
still somehow depends on the endpoint behavior of the light-meson's
distribution amplitude.

Recently, a QCD factorization formula \cite{s6}
is proposed for $B\rightarrow \pi \pi$, $\pi$K and $\pi$D. It makes
a great progress to deal with nonleptonic decays of B meson. In this
approach, the amplitudes for these decays are expressed in terms of the
semileptonic form factors, hadronic light-cone distribution amplitudes and
hard-scattering functions that are calculable in PQCD, and the
semileptonic form factors and the distribution amplitudes are taken as
inputs since the
form factors can be measured experimentally and the distribution
amplitudes are universal function of the single meson state.
Theoretically, the precise calculations of heavy-to-light form factors
are of a great interest. Especially, it will be helpful to a clear 
understanding of
$B\rightarrow \pi \ell \widetilde{\nu_ \ell}$ ($\ell=e, \mu$) which
provides us with
a good chance to extract the Cabibbo-Kobayashi-Maskawa (CKM) matrix
element $\mid{V_{ub}}\mid$ from the available data.

The fact that a
considerable long-distance contribution may dominate the heavy-light 
form factor has been a motivation for applying the light-cone QCD sum 
rules to the $B\rightarrow\pi$ weak form factor \cite{s7}. In this 
approach, the non-perturbative dynamics is parametrized as so-called
light-cone wave functions classified by their twist.
Remarkably, the main uncertainties in the sum rule results arise from
light-cone wave functions. Now only the twist-2
wave functions, which is of dominate contributions to the sum rules, has
systematically been investigated. This is not the
case, however, for the twist-3 and the twist-4 wave functions, which are
understood poorly. On the other hand, although QCD radiative
corrections to the twist-2 term are
considered in \cite{s8}, for improving the predictions, and their 
impact on the sum rule is found out to be negligible small, numerical
results are less convincing, because we have no reason to believe that
$O(\alpha _s)$ corrections to the twist-3 terms can safely be
neglected. From the above analyses, we can conclude that
the great uncertainty, if possible, would be due to the uncertainties in
the twist-3 wave functions and the lack of 
the corresponding $O(\alpha_s)$ corrections, in the existing calculations
of $B\rightarrow \pi $ form factors
in the framework of the light-cone sum rules.

In the present work, we suggest an improved  approach to calculating
heavy-to-light weak form factors, and then apply it to re-analyze
$B\rightarrow \pi \ell \widetilde{\nu _\ell }$. The striking advantage of
the method is, as will be
shown in the following, that contributions of the twist-3 wave functions
vanish at all from the light-cone sum rule in question, such that the
possible pollution by them is effectively avoided. It will be 
beneficial to enhancing the reliability of the light-cone sum rule
calculations.
\section * {II. CORRELATOR}
 
Let us start with the following definition of $B\rightarrow \pi $ weak
form factors $f(q^2)$ and $\widetilde{f}(q^2)$:
\begin{equation}
\langle\pi (p)|\overline{u}\gamma _\mu b|B(p+q)\rangle=2f(q^2)p_\mu
+\widetilde{f} (q^2)q_\mu ,
\end{equation}
with $q$ being the momentum transfer. Following Refs. \cite{s9}, we choose
to use a chiral current
\begin{eqnarray}
\Pi_\mu (p,q) && =i\int d^4xe^{iqx}\langle\pi (p)|T
\{ \overline{u}(x)\gamma _\mu (1+\gamma
_5)b(x),\overline{b}(0)i(1+\gamma _5)d(0)\}|0\rangle \nonumber \\ &&
=\Pi (q^2,(p+q)^2)p_\mu +\widetilde{\Pi }(q^2,(p+q)^2)q_\mu ,
\end{eqnarray}
which is different from that in Ref. \cite{s3} and \cite{s4} to calculate
$f(q^2)$ and $\tilde{f}(q^2)$.
Here the T product of the chiral current operator is inserted between the
vacuum and the on-shell $\pi$ meson state.

First, we discuss the hadronic representation for the
correlator. This can be done by inserting the complete intermediate states
with the same quantum numbers as the current operator
$\bar bi(1+\gamma _5)d$ in the correlator. By isolating the pole
term of the lowest pseudoscalar B meson, we have the hadronic 
representation in the following
\begin{eqnarray}
\Pi_\mu ^H(p,q)  &&= \Pi ^H(q^2,(p+q)^2)p_\mu +
\widetilde{\Pi}^H(q^2,(p+q)^2)q_\mu
 \nonumber \\
&&= \frac{ \langle \pi|\overline{u}\gamma _\mu b|B\rangle \langle
B|\overline{b}\gamma_5d|0\rangle} {m_B^2-(p+q)^2} \nonumber \\
&&+\sum\limits_H\frac{\langle \pi |\overline{u}\gamma _\mu
(1+\gamma _5)|B^H\rangle \langle B^H|\overline{b}i(1+\gamma
_5)d|0\rangle}{m_{B^H}^2-(p+q)^2}.
\end{eqnarray}

Note that the intermediate states $B^H$ contain not only
pseudoscalar resonances of the masses greater than $m_B$, but also
scalar resonances with $J^p=0^{+}$, corresponding to the operator
$\bar bd$. With Eq. (1) and the definition $\langle B|\bar{b}i\gamma
_5d|0\rangle={m_B}^2f_B/m_b$, the invariant amplitudes $\Pi^H$ and
$\tilde{\Pi}^H$ read off

\begin{equation}
\Pi^H(q^2,(p+q)^2)=\frac{2 f(q^2) m_B^2 f_B}{m_b
(m_B^2-(p+q)^2)} +\int \limits_{s_0}^{\infty}\frac
{\rho^H(s)}{s-(p+q)^2}ds +subtractions,
\end{equation}

and

\begin{equation}
\widetilde{\Pi} ^H(q^2,(p+q)^2)=\frac{\widetilde{f}(q^2) m_B^2
f_B}{m_b
(m_B^2 -(p+q)^2)} +\int\limits_ {s_0}^{\infty}\frac {\widetilde{\rho}%
^H(s)}{s-(p+q)^2}ds +subtractions,
\end{equation}
where we have replaced the contributions of higher resonances and continuum
states with dispersion integrations, in which the threshold parameter
$s_0$
should be set near the squared mass of the lowest scalar $B$ meson, and
the spectral densities $\rho^H(s)$ and $\tilde{\rho}^H(s)$ can be
approximated by
invoking the quark-hadron duality ansatz
\begin{equation}
\rho ^H(s)(\widetilde{\rho }^H(s))=\rho ^{QCD}(s)(\widetilde{\rho }%
^{QCD}(s))\theta (s-s_0).
\end{equation}
If we confine ourselves to discussing the semileptonic decays $B\rightarrow
\pi \ell \tilde {\nu_\ell }$ ($\ell =e,\mu $), the contributions of $%
\widetilde{f}(q^2)$ to the decay amplitudes are small enough to be
negligible, due to the smallness of the final state lepton masses, and
therefore only the form factor $f(q^2)$ need considering.

On the other hand, we have to calculate the corrector in QCD theory, to
obtain the desired sum rule for $f(q^2)$. It is possible by using the
light
cone OPE method. To this end, we work in the large space-like momentum
regions $(p+q)^2-m_b^2\ll 0$ for the $b\bar d$ channel, and $q^2\ll
m_b^2-O(1GeV^2)$ for the momentum transfer, which correspond to the small
light cone distance $x^2\approx 0$ and are required by the validity of the
OPE. In addition, the chiral limit $p^2=m_\pi ^2=0$ is taken throughout this
discussion, for simplicity. The leading contribution to the OPE is easy to
drive by contracting the $b$-quark operates to a free propagator. After
further considering the effect of the background gluon field, we can write
down a full $b$-quark propagator

\begin{eqnarray}
\langle0|T{b(x)\bar{b}(0)}|0\rangle &&
=iS_b^{(0)}(x)-ig_s\int \frac{d^4k}{(2\pi
)^4}e^{-ikx}\int\limits_0^1dv \left[\frac {1}{2}\frac{\hat{
k}+m}{(m_b^2-k^2)^2}G^{\mu \nu }(vx)\sigma _{\mu \nu }  \nonumber
\right. \\
&& \left. +\frac 1{m_b^2-k^2}vx_\mu G^{\mu \nu }(vx)\gamma _\nu \right].
\end{eqnarray}
Here $G_{\mu \nu }$ is the gluonic field strength, $g_s$ denotes the strong
coupling constant and $S_b^0(x)$ expresses a free $b$-quark
propagator
\begin{equation}
S_b^{(0)}(x)=\int \frac{d^4k}{(2\pi )^4}e^{-ikx}\frac{\hat
{k}+m}{k^2-m_b^2}.
\end{equation}

Consider first the leading contribution from the free $b$-quark
propagator. Carrying out the OPE for the corrector and making use of the
Eq. (8), we have
\begin{equation}
\Pi ^{(\bar{q}q)}=-2m_bi\int \frac{d^4xd^4k}{(2\pi
)^4}e^{i(q-k)x}\frac 1{k^2-m_b^2}\langle\pi (p)|T\bar{u}(x)\gamma
_\mu \gamma _5d(0)|0\rangle,
\end{equation}
for the two-particle contribution $\Pi^{(\bar{q}q)}$. An
important observation, as have been emphasized, is that only the
leading nonlocal matrix element 
$\langle\pi (p)|T\bar{u}(x)\gamma_\mu \gamma _5d(0)|0\rangle$
contributions to the corrector, while the nonlocal matrix elements
$\langle\pi (p)| \bar{u}(x)i\gamma _5d(0)|0\rangle$ and $\langle\pi
(p)|\bar{u}(x)\sigma _{\mu \nu }\gamma _5d(0)|0\rangle$
whose leading terms are of twist-3, disappear in
our approach. Proceeding to Eq. (9), we discuss the light cone expansion 
of $ \langle\pi (p)|T\bar{u}(x)\gamma _\mu \gamma
_5d(0)|0\rangle$. In general, for a nonlocal quark -antiquark
operator we expand it around $x=0$, and then parametrize the
operator matrix elements of any definitive twist by the so-called
light-cone
wave functions. In the present case, the nonlocal matrix element 
$\langle\pi (p)|T%
\overline{u}(x)\gamma _\mu \gamma _5d(0)|0\rangle$ can be expanded
as
\begin{eqnarray}
\langle\pi (p)|T\bar{u}(x)\gamma _\mu \gamma _5d(0)|0\rangle &&
=-ip_\mu f_\pi \int\limits_0^1due^{iupx}(\varphi _\pi
(u)+x^2g_1(u)) \nonumber \\ && +f_\pi (x_\mu -\frac{x^2p_\mu
}{px})\int\limits_0^1due^{iupx}g_2(u),
\end{eqnarray}
to the twist-4 accuracy. Where $\varphi_\pi (u)$ is the twist-2 wave 
function, 
while both $g_1(u)$ and $g_2(u)$ have twist-4. Substituting Eq. (10) into
Eq. (9) and
integrating over $x$ and $k$ yields
\begin{eqnarray}
\Pi ^{(\bar{q}q)}(q^2,(p+q)^2) && =2f_\pi m_b \left[\int\limits_{0}^{1}
\frac{du}{u}\varphi _{\pi}(u)\frac
{1}{s-(p+q)^2}-8m_b^2\int\limits_{0}^{1}\frac{du}{%
u^3}g_1(u)\frac{1}{(s-(p+q)^2)^3} \nonumber \right. \\
&& \left. +2\int\limits_{0}^{1}\frac{du}{u^2}%
G_2(u)\frac{1}{(s-(p+q)^2)^2}+4\int\limits_{0}^{1}\frac{du}{u^3}G_2(u)\frac{%
q^2+m_b^2}{(s-(p+q)^2)^3} \right],
\end{eqnarray}
with $G_2(u)=\int \limits_{0}^{u} g_2(v) dv$. In deriving Eq. (11) the 
relation $u=\frac{m_b^2-q^2}{s-q^2}$ has been used, and thus it should be 
understood
that $s$ is the function of argument $u$. A further discussion involves
the
evaluations of higher Fock-state effects. This can be done by taking into
account the second term in Eq. (7) in the OPE of the correlator. A
straightforward calculation gives for the three-particle contribution 
$\Pi_\mu ^{(\bar{q}qg)}$

\begin{eqnarray}
\Pi_\mu^{(\bar{q}qg)}(q^2,(p+q)^2) && = i g_s m_b \int
\frac{d^4kd^4xdv}{(2 \pi)^4 (m_b^2-k^2)} e^{i (q-k) x}
(\langle\pi(p)|\bar{d}(x)\gamma_\mu G^{\alpha
\beta}(v x) \sigma_{\alpha \beta}u(0)|0\rangle \nonumber \\ 
&& + \langle\pi(p)|\bar{d}%
(x)\gamma_\mu\gamma_5G^{\alpha \beta}(v x) \sigma_{\alpha
\beta}u(0)|0\rangle ).
\end{eqnarray}
Considering $\langle\pi(p)|\bar{d}(x)\gamma_\mu G^{\alpha
\beta}(v x) \sigma_{\alpha \beta}u(0)|0\rangle=0 $, as required by the
parity conservation in strong interaction, and using the identity

\begin{equation}
\gamma_\mu \sigma_{\alpha \beta}=i (g_{\mu \alpha} \gamma_\beta-
g_{\mu\beta}\gamma_\alpha)+\epsilon_{\mu \alpha \beta\nu}\gamma^\nu\gamma_5.
\end{equation}
We further have
\begin{eqnarray}
\Pi_\mu^{(\bar{q}qg)}(q^2,(p+q)^2) && =im_b\int \frac{d^4kd^4xdv} {(2 \pi)^4
(m_b^2-k^2)} e^{i (q-k) x} \left(i g_{\mu \alpha} \langle\pi(p)|\bar{u}(x)
\gamma_\beta\gamma_5 g_s G^{\alpha \beta}(v x)d(0)|0\rangle \nonumber 
\right. \\ 
&& \left. +\langle\pi(p)|\bar{u}(x)\gamma^{\nu} g_s\widetilde{G}_{\mu
\nu}(vx)d(0)|0\rangle \right),
\end{eqnarray}
with $\widetilde{G}_{\mu \nu}(v x)=\frac{1} {2}\epsilon^{\mu \nu \sigma
\tau} G^{\sigma \tau}(v x)$ it
should be noted that situation here is all the same as that in
Eq. (11): the nonlocal matrix
element
$\langle\pi|\overline{u}(x)\sigma_{\mu\nu}\gamma_5g_sG_{\alpha \beta}
(vx)d(0)|0\rangle$, which has the twist-3 in leading-order in the
light-cone expansion, vanishes from the OPE. As a result, a
self-consistency
is kept in our approach. The matrix elements in
Eq. (14) can be parametrized in terms of
the three -particle wave functions of twist-4 $\varphi_\perp$, $%
\varphi_\parallel$, $\tilde{\varphi_\perp}$ and $\tilde{\varphi_\parallel}$
defined by
\begin{eqnarray}
\langle\pi(p)| \bar{d}(x)\gamma_\mu\gamma_5g_sG_{\alpha \beta}(v
x)u(0)|0\rangle && =f_\pi[q_\beta(g_{\alpha \mu}-\frac{x_\alpha
q_\mu}{q x})-q_\alpha (g_{\beta \mu}-\frac{x_\beta q_\mu}{q x})]  
 \int D\alpha_i\varphi_\perp(\alpha _i) e^{i q x (\alpha_1+v
\alpha_3)} \nonumber \\ && + f_\pi \frac{q_\mu}{q x} (q_\alpha x_\beta-q_ 
\beta x_\alpha) \int D\alpha_i\varphi_\parallel(\alpha _i) e^{ i q x
(\alpha_1+v \alpha_3)},
\end{eqnarray}
\begin{eqnarray}
\langle\pi(p)| \bar{d}(x)\gamma_\mu g_s\widetilde{G}_{\alpha
\beta}(v x)u(0)|0\rangle && =i f_\pi[q_\beta(g_{\alpha
\mu}-\frac{x_\alpha q_\mu}{q x})-q_\alpha (g_{\beta
\mu}-\frac{x_\beta q_\mu}{q x})] \int
D\alpha_i\widetilde\varphi_\perp(\alpha_i) e ^{i q x (\alpha_1+v
\alpha_3)} \nonumber \\ && + i f_\pi \frac{q_\mu}{q x} (q_\alpha 
x_\beta-q_\beta
x_\alpha) \int D\alpha_i\widetilde\varphi_\parallel(\alpha_i) e^{i
q x (\alpha_1+v \alpha_3)},
\end{eqnarray}
with $D\alpha_i=d\alpha_1 d\alpha_2 d\alpha_3 \delta(1-\alpha_1-\alpha_2
-\alpha_3)$. Completing the integrations over $x$ and $k$, we have
\begin{equation}
\Pi^{(\bar{q}qg)}(q^2,(p+q)^2) = 2 m_b f_\pi \int \limits_{0}^{1} dv \int
D\alpha_i \frac {2 \varphi_{\perp}(\alpha_i)
+2\tilde{\varphi}_{\perp}(\alpha_i)
-\varphi_{\parallel}(\alpha_i)-\tilde{\varphi}_{\parallel}(\alpha_i)}
{[s-(p+q)^2]^2 (\alpha_1+v \alpha_3)^2},
\end{equation}
with parameter $s$ defined by the relation $\alpha_1+v \alpha_3=\frac{%
m_b^2-q^2}{s-q^2}$. 
The final light-cone QCD expansion of the correlator can
be written down as
\begin{equation}
\Pi^{QCD}(q,(p+q))=\Pi^{(\bar{q}q)}(q,(p+q))+\Pi^{(\bar{q}qg)}(q,(p+q))
\end{equation}
\section * {III. SUM RULE FOR $f(q^2)$}

Further, to carry out the subtraction procedure of the continuum spectrum 
we need to
convert the QCD representation (18) into a dispersion integration. In $%
\Pi^{QCD}$, the term proportional to $\frac{1}{s-(p+q)^2}$ in integrand
is already a dispersion integration with respect to $(p+q)^2$ so that
subtraction of the continuum can be made by simply changing the lower limit
of integration from 0 to $\Delta=\frac{m_b^2-q^2}{s_0-q^2}$, while those with
higher power of $\frac{1}{s-(p+q)^2}$, after the partial integration, become
the following form
\begin{equation}
I=\int \limits_{m_b^2}^{\infty}\frac{F(s)}{s-(p+q)^2} ds,
\end{equation}
which being a dispersion integration with the perturbative spectrum densities
$F(s)$. For instance, we have $F(s)=\frac {d^{2} f(s)}{d s^{2}}$, with
$f(s) =8f_\pi m_b^3 g_1(u(s)) \frac{q^2-s}{(m_b^2-q^2)^2}$, for the
contribution of
the twist-4 wave function $g_1(u)$ in Eq. (11). In this case, the 
subtraction of the continuum corresponds to a simple replacement 
$\infty\rightarrow s_0$.

Now, the light cone QCD sum rule for $f(q^2)$ can be obtained, by making
the Borel transformations with respect to $(p+q)^2$ in the hadronic and the
QCD expressions and equating them. The result is
\begin{eqnarray}
f(q^2) && = \frac{m_b^2 f_\pi}{m_B^2 f_B} e^{\frac{m_B^2}{M^2}}
\left\{\int \limits_{\triangle}^{1}\frac{du}{u} e^{-\frac{m_b^2-q^2
(1-u)}{u M^2}} \left(\varphi_ \pi(u)-\frac{4 m_b^2}{u^2 M^4} g_1(u) +
\frac{2}{u M^2} \int \limits_{0}^{u}g_2(v) dv
(1+\frac{m_b^2+q^2}{u M^2})\right) \nonumber \right. \\ && +\int
\limits_{0}^{1}
dv \int
D\alpha_i\frac {\theta(\alpha_1+v \alpha_3 -\Delta)}{(\alpha_1+v
\alpha_3)^2 M^2} e^{-\frac {m_b^2-(1-\alpha_1- v \alpha_3)
q^2}{M^2 (\alpha_1+v \alpha_3)}} (2 \varphi_\perp(\alpha_i)+2
\widetilde\varphi_i\perp(\alpha_i)
-\varphi_\parallel(\alpha_i)-\widetilde\varphi_\parallel(\alpha_i))
\nonumber \\ && -4 m_b^2 e^{ \frac{-s_0}{M^2}}
\left(\frac{1}{(m_b^2-q^2)^2}
(1+\frac{s_0-q^2} {M^2}) g_1(\Delta)-\frac{1}{(s_0-q^2)
(m_b^2-q^2)} \frac {dg_1(\Delta)} {du} \right) \nonumber \\ &&
\left. -2e^{\frac{-s_0}{M^2}} \left(\frac{m_b^2+q^2}{(s_0-q^2) (m_b^2-q^2)}
g_2(\Delta) - \frac{1}{(m_b^2-q^2)} (1+\frac{m_b^2+q^2}{m_b^2-q^2}
(1+\frac{s_0-q^2}
{M^2}) \int \limits_{0}^{\Delta}g_2(v) dv \right) \right\}.
\end{eqnarray}
We would like to stress that the terms proportional to exponential
factor $e^{\frac{-s_0}{M^2}}$ arise from the substructions of the continuum,
and may not be neglected for our present purposes.

Before proceeding further
we need to make a choice of input parameters entering the sum rule for $%
f(q^2)$. To begin with, let us specify the set of pion wave functions. 
For the leading twist-2 wave function $\varphi_\pi(u)$, the asymptotic 
form is exactly
given by PQCD \cite{s10} $\varphi_\pi(u,\mu \rightarrow 
\infty)=6u(1-u)$, nonperturbative corrections can be included in a 
systematic way in term of the approximate conformal invariance of QCD 
\begin{equation}
\varphi_\pi(u,\mu)=6 u (1-u) [1+a_2(\mu) C_2^{\frac{3}{2}}(2 u-1)+ a_4(\mu)
C_4^{\frac{3}{2}} (2 u-1)+\cdots],
\end{equation}
with the Gegenbaer Polynomials
\begin{equation}
C_2^{\frac{3}{2}}(2 u-1)=\frac{3}{2} [5 (2 u-1)^2-1],
\end{equation}
\begin{equation}
C_4^{\frac{3}{2}}(2 u-1)=\frac{15}{8} [21 (2 u-1)^4-14 (2 u-1)^2+1].
\end{equation}
The coefficients in the expansion $a_n(\mu)$ can be determined by a certain
nonperturbative approach. As we know, there are many models for the 
twist-2 wave function \cite{s11}. In order to make a comparison with 
the previous result, we follow the Ref. \cite{s12} and use
\begin{equation}
a_2(\mu_0=0.5GeV)=\frac{2}{3}, a_4(\mu_0=0.5GeV)=0.43,
\end{equation}
which result from a analysis of light-cone sum rules for the $\pi N N$ and
the $\omega \rho \pi$ couplings. Furthermore, the use of the 
Renormalization Group Equation (RGE) gets
\begin{equation}
a_2(\mu_b)=0.35, a_4(\mu_b)=0.18,
\end{equation}
at the scale $\mu_b=\sqrt{m_B^2-m_b^2} \approx 2.5 GeV$, which characterizes
the mean virtuality of the $b$ quark. For the twist-4 wave functions, we
use
the results for the three-particle wave functions \cite{s12}
\begin{eqnarray}
&& \varphi_{\perp}(\alpha_i)=30 \delta^2 (\alpha_1-\alpha_2) \alpha_3^2
[\frac{1}{3}+2 \epsilon (1-2 \alpha_3)] ,
\nonumber \\ && \widetilde{\varphi}_{\perp}(\alpha_i)=30
\delta^2 \alpha_3^2 (1-\alpha_3) [\frac{1}{3}+2 \epsilon (1-2\alpha_3)] ,
\nonumber \\ && \varphi_{\parallel}(\alpha_i)=120 \delta^2 \epsilon 
(\alpha_1-\alpha_2)
\alpha_1 \alpha_2 \alpha_3 ,
\nonumber \\ && \tilde{\varphi}_{\parallel}(\alpha_i)=-120
\delta^2 \alpha_1 \alpha_2 \alpha_3 [\frac{1}{3}+\epsilon (1-3\alpha_3)],
\end{eqnarray}
with $\delta^2(\mu_b)=0.17GeV^2$ and $\varepsilon(\mu_b)=0.36$. Further, a
relation can be obtained between the two-particle twist-4 wave functions and
the above these by equation of motion such that we have \cite{s12}
\begin{eqnarray}
g_1(u) && =
\frac{5}{2} \varepsilon^2 u^2 \overline{u}^2+\frac{1}{2} \varepsilon 
\delta^2 
\nonumber \\ && [u \overline{u} (2+13 u \overline{u})+10 u^3 \ln u (2-3 
u+\frac{6}{5} u^2) \nonumber \\ &&
+10 \overline{u}^3\ln{\overline{u}}(2-3\overline{u}+\frac{6}{5}
 \overline{u}^2)] , \nonumber \\ g_2(u) && =\frac{10}{3}\delta^2
u\overline{u}(u-\overline{u}).
\end{eqnarray}
Unlike the case of the twist-2 wave functions, these twist-4 wave functions
seem to be very difficult to test by experiment, for they usually are of
negligible contributions in the sum rules.

Another important input is the decay constant of B meson $f_B$. The
QCD sum rule for $f_B$ has been discussed many times. However, all these
estimates are not applicable in our sum rule for $f(q^2)$. The reason is
that in the present case a chiral current correlator is adopted to avoid
pollution by the twist-3 wave functions, so that a similar correlator
has to be used, for consistency, in the sum rule calculation of $f_B$. To 
this end, we consider the following two-point correlator 
\begin{equation}
K(q^2)=i\int d^4xe^{iqx}\langle0|\overline{q}(x)(1+\gamma
_5)b(x),\overline{b}(0) (1-\gamma_5)q(0)|0\rangle.
\end{equation}

The calculation should be limited to leading order in QCD, since the QCD
radiative corrections to the sum rule for $f(q^2)$ are neglected as well. A
standard manipulation yields three self-consistent sets of 
results\cite{s9}: (1) $f_B=165 MeV$, for $m_b=4.7GeV$ and $s_0=33GeV^2$,
(2)$f_B=120MeV$, for $m_b=4.8GeV$ and $s_0=32GeV^2$, and (3) $f_B=85MeV$,
for $m_b=4.9GeV$ and $s_0=30GeV^2$. The above results correspond to the
best fit in $s_0$ and will
be used as inputs in numerical analyses of the sum rule for $f(q^2)$. At
the point, a few comments are in order: (1) some vacuum condensate
parameters vanish from the sum rule for $f_B$, and thus some inherent
uncertainties in the sum rule are reduced. (2) the threshold parameters
$s_0$ turn out to be of values less
than those in the conventional sum rule for $f_B$. This is consistent 
with the case in the sum rule for $f(q^2)$. As for the B meson mass $m_B$
and the pion decay constant $f_\pi$, we
take the present world average value $m_B=5.279GeV$, and $f_\pi=0.132GeV$.
\section * {IV. NUMERICAL RESULT}

With these inputs, we can carry out the numerical analysis. The first step
is, according to the standard procedure, to look for a range of the Borel
parameter $M^2$, in which the numerical results are quite stable for a
given
threshold $s_0$. Then, what remains to be done is to determine the fiducial
interval of $M^2$, from which the desired sum rule results can be read 
off, by the requirement that the contributions of the twist-4
wave functions do not exceed $10\%$, while those of the continuum states are
not more than $30\%$.

In the present case, the reasonable range of $M^2$, for the threshold $s_0$
given above, is found to be $8GeV^2 \leq M^2 \leq 17GeV^2$ with the 
different central values as $q^2$ changes. In such a "window", $f(q^2)$
depends very weakly on $M^2$, up to $q^2= 18GeV^2$. As it is shown, for
example, in Fig. 1 where the two typical cases, corresponding to 
$q^2=10 GeV^2$ and $16GeV^2$, are considered for an illustrative
purpose. This allows us to estimate safely
the variation of $f(q^2)$ with $q^2$, at a certain specific value of
$M^2$. The numerical results at $M^2=12GeV^2$, together 
with the previous light-cone
sum rule prediction \cite{s7} are plotted in Fig. 2, for a comparison. We 
find $f(0)=0.27$, $0.29$, and $0.33$ (corresponding to set(3), set(2), and
set(1), respectively), which are in basic agreement with the result in 
\cite{s7} $f(0)=0.29$. As a mater of fact, numerical agreement between 
the two
different approaches exists up to $q^2=10GeV^2$, the differences being 
within $20\%$. The obvious numerical derivation, however, begins to appear 
beyond $10GeV^2$ and our results turn out to be less than those 
of \cite{s7} by about $(35-40)\%$, near $q^2=18GeV^2$. Apparently, the
fact that $f(q^2)$ is less sensitive to $M^2$ can not account for the
disagreement. To clarify this issue, both approaches 
have to undergo a more systematic investigation, including a complete
evaluation of $O(\alpha_{s})$ corrections and a detailed
analysis of the uncertainties in the twist-3 wavefunctions. Indeed,
the radiative corrections
as it has been shown in \cite{s8} are negligibly small for the
twist-2 term. It perhaps is not the case in our approach and for the
twist-3 terms in the sum rules of \cite{s7}. In the region 
$q^2\geq18GeV^2$,
applicability of the light-cone sum rules is questionable, as has been
mentioned, such that a comparison between the different approaches is
meaningless.

It is needed to make a systematic discussion on the sources of 
uncertainties for $f(q^2)$. All the above calculations correspond
to taking the central values of threshold parameters, which are determined in
the two-point sum rule for $f_B$. To look at the numerical impact of the
uncertainties in threshold parameter on the sum rule for $f(q^2)$, we make
use of the analytic form, instead of the numerical results, for
the two point sum rule for $f_B$ in the numerical calculations. It is shown 
that the resulting $f(q^2)$ varies by $(10-15)\%$ relative to the central 
values, depending on $m_b$ and $q^2$. Also, we investigate the sensitivity 
of $f(q^2)$ to the simultaneous
variations of $s_0$ and $m_b$ in the regions $30GeV^2 \leq s_0 \leq 33GeV^2$
and $4.7GeV \leq m_b \leq 4.9GeV$, finding that the induced change in $f(q^2)$
in the case is less than $5\%$ in the total range of $q^2$, for the most
stable values of $f_B$, and therefore is negligible.

In addition, there also are the uncertainties related to the light-cone
wave functions of $\pi$ meson. For example, the wave function, which
is closed to the asymptotic form, will give smaller value of $f(0)$.
However, the twist-2 wave function is universal for the different
processes. The uncertainties due to it can be controlled well as
soon as one can obtain more reliable twist-2 wave function to fit
them. For the twist-4
wave functions, considering that they have only the effect of
about $(4-6)\%$ on $f(q^2)$, as shown, we can imagine that the contributions of 
wave functions beyond the twist-4 are anyway negligibly small. In 
fact, this signals that we need
not be careful about the sensitivity of $f(q^2)$ to wave functions of
twist-4 and beyond twist-4.
As the twist-3 wave functions as go, the numerical calculations show that
their contributions are comparable with those of the twist-2, amounting to
about $50\%$ in \cite{s7}. Remarkably, the reliability of these wave 
functions
has to be subject to a test in that case. Nevertheless, this causes no 
problem in the present case, for all the twist-3 wave functions make a 
vanishing contribution to the sum rule in question, up to all orders in 
PQCD.
\section * {V. SUMMARY}

To summarize, we have re-examined that weak form factor $%
f(q^2)$ for B decays into light pseudoscalar mesons, taking $B\rightarrow 
\pi$ semileptonic transitions as an illustrative example, in the 
light-cone QCD sum rule framework. The aim is to control the nonperturbative
dynamics in the sum rules, to the best of our ability, and further to 
enhance the predictivity and reliability of numerical results.
To this end, a chiral-current correlator is worked out. It is explicitly 
shown that the twist-3 light-cone wave functions, which have not been
understood very well, can be effectively eliminated from the our sum rule
for $f(q^2)$. Consequently, the possible pollution by them is avoided in 
the final expression.  The results presented here will be beneficial to
the precision extracting of the
CKM matrix element $|V_{ub}|$ from the exclusive processes $%
B\rightarrow\pi\ell\tilde{\nu_l}$ $(l=e,\mu)$, by confronting the theoretical
predictions with the experimentally available data.

In comparison to a previous estimate based on the light-cone sum rules, we
find that the numerical agreement exists
between the two different sum rules for $f(q^2)$  in the 
region of
momentum transfer $0 \leq q^2 \leq 10GeV^2$; beyond this region, a
remarkable
numerical deviation begins to appear; in particular, near $q^2=18GeV^2$,
( maximum value required by the light-cone OPE ), our numerical results are
less than that in \cite{s7} by about $(35-40)\%$. Also, the
possible uncertainties in the sum rule $f(q^2)$ due to the parameter $m_b$ 
are discussed. At present we haven't included the 
PQCD radiative corrections. It is expected that our result doesn't
change much after including the PQCD radiative corrections since the
twist-3 light-cone wave functions are eliminated at all in our approach. 
\section * {Acknowledgements}

One of the authors X. Y. Wu thanks L. Z. Hai, Y. D. Shan and G. H. Zhu
for useful comments.

\newpage

\begin{figure}[tb] 
\vspace*{2cm}
\centerline{\epsfig{figure=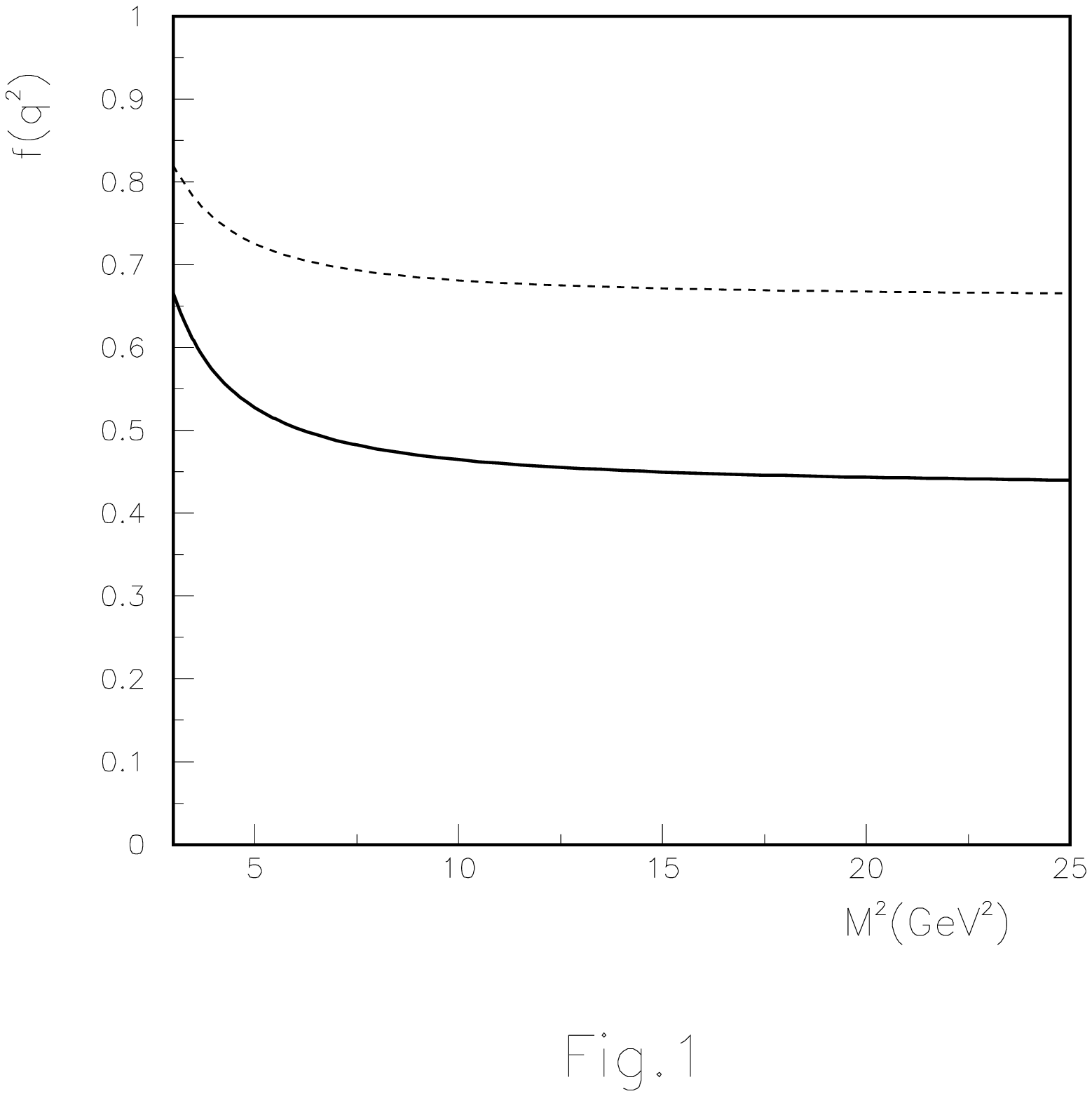, height=12cm, width=16cm, angle=0}}
\vspace*{1.cm} 
\caption{Sensitivity of the form factor $f(q^2)$ to the Borel parameter
$M^2$. Considered are the two typical cases of $q^2=10GeV^2$ (solid) and
$q^2=16GeV^2$ (dashed), with $S_{0}=32GeV^2$ and $m_{b}=4.8GeV$.}
\end{figure}

\begin{figure}[tb] 
\vspace*{2cm}
\centerline{\epsfig{figure=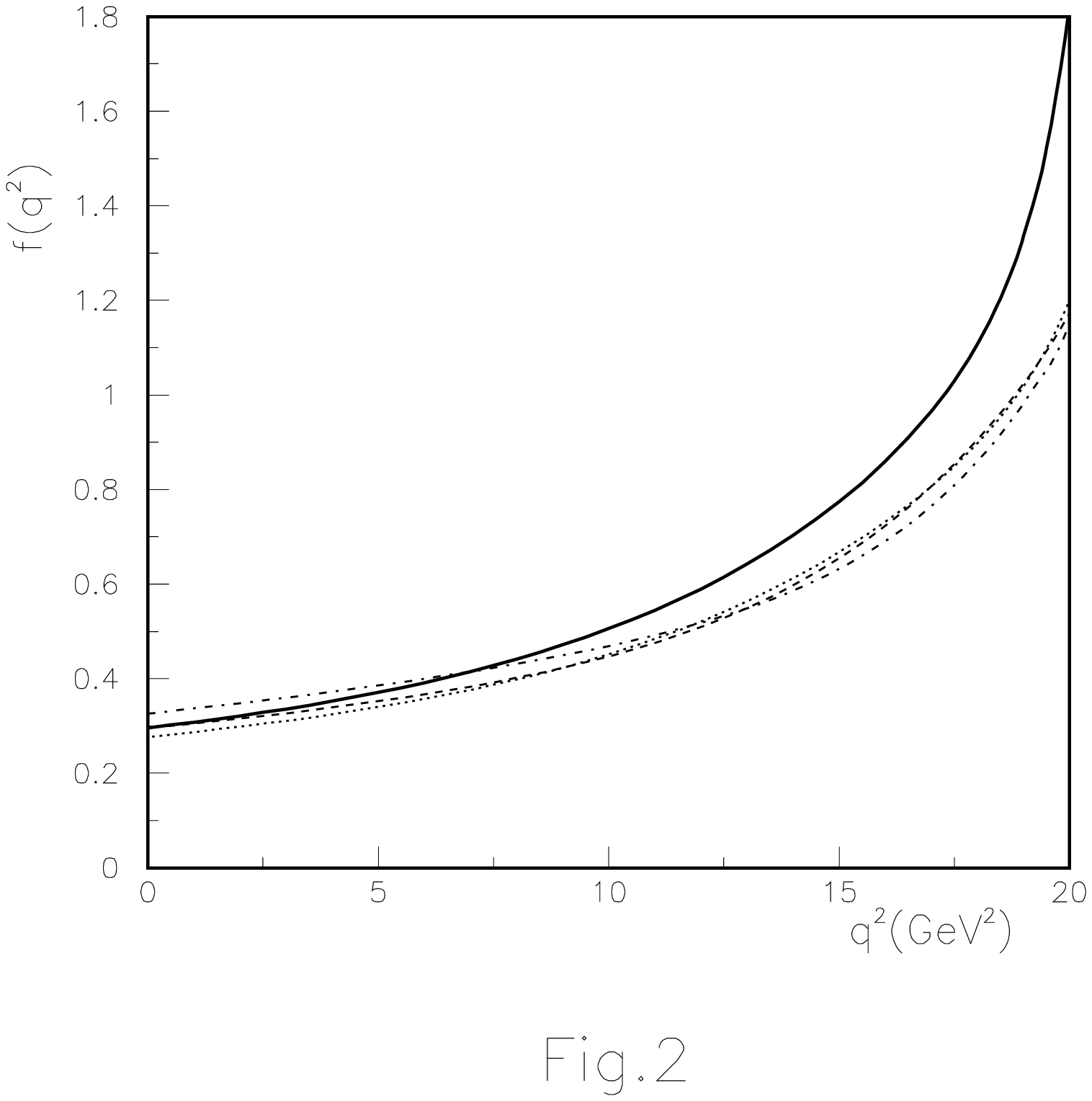, height=12cm, width=16cm, angle=0}}
\vspace*{1.cm} 
\caption{The light-cone QCD sum rules for form factor $f(q^2)$ of 
$B\rightarrow \pi$ semileptonic transitions at $M^2=12GeV^2$. The solid 
curve expresses the results in [7], while The dotted, the dashed and
the dashed-dot curves correspond to our predictions, with 
(i)$m_b=4.7GeV$, $s_0=33GeV^2$, (ii) $m_b=4.8GeV$, $s_0=32GeV^2$ and 
(iii) $m_b=4.9GeV$, $s_0=30GeV^2$, respectively.}
\end{figure}

\end{document}